\documentclass[sigplan,10pt]{acmart}
\usepackage{wasysym}

\renewcommand\footnotetextcopyrightpermission[1]{}   

\title{Intrinsic Value Modules: Trusted-Hardware Support for Locally Final Offline Asset Transfer}

\author{AmirReza Fathi}
\affiliation{%
  \institution{Tabdeal Labs}
  \city{Tehran}
  \country{Iran}
}
\email{amirreza.fathi@tabdeal.org}

\author{Ali Rahimi}
\affiliation{%
  \institution{Tabdeal Labs}
  \city{Tehran}
  \country{Iran}
}
\email{ali.rahimi@tabdeal.org}
\authornote{Corresponding author.}

\begin{document}

\pagestyle{plain}


\begin{abstract}
Consensus lets distributed participants agree on a common state, but it makes monetary validity depend on network connectivity and collective acceptance. We explore making validity intrinsic to a digital coin. We propose an \emph{Intrinsic Value Module (IVM)}: trusted hardware that gives a coin locally verifiable authenticity and spendability, analogous to verifying a signature. Under stated hardware and cryptographic assumptions, the IVM abstraction and our offline transfer protocol move monetary uniqueness from a replicated ledger into non-clonable protected state. The protocol prevents forgery, replay, double credit, and double spending by genuine, uncompromised IVMs; communication failures may lock value but cannot duplicate it. We distinguish local transfer atomicity and asset uniqueness, which the protocol establishes, from global state consistency across disconnected devices, which it does not. The IVM does not eliminate trust; it relocates trust to certified hardware, issuer roots, and non-rollback state. We present the architecture, protocol, and security analysis, and discuss open challenges.
\end{abstract}

\maketitle


\keywords{
  Distributed systems,
  state management,
  consensus,
  blockchain
}


\section{Introduction}

Consensus has long been a central problem in distributed computing: participants
must establish a common view of state even when communication is asynchronous,
machines fail, or some participants behave maliciously. Classical work exposed
both the limits of agreement in asynchronous systems~\cite{fischer1985impossibility}
and practical ways to obtain it, from Paxos~\cite{lamport1998parttime} to
Byzantine-fault-tolerant replication~\cite{castro1999practical}. Bitcoin's
proof-of-work protocol transformed these ideas by enabling open, mutually
distrustful participants to agree on an ordered payment history without a
central operator~\cite{nakamoto2008bitcoin}; subsequent blockchain systems made
consensus a foundation for programmable assets and applications. This
development has had significance beyond distributed-systems research: it
created new markets for digital assets, enabled permissionless financial
coordination across organizational and national boundaries, and broadened the
design space for economic institutions and international exchange
~\cite{catalini2020some,werbach2018blockchain}.

That achievement comes with an important dependency. Although a user can
verify a signature locally, a Bitcoin payment becomes durable only after miners
include it and extend the corresponding chain, and an Ethereum transaction
depends on validators agreeing on the canonical state. Thus, ownership and
spendability are not intrinsic to a coin: they depend on other participants,
network connectivity, and the history those participants accept. Temporary
forks and chain reorganizations can reverse transactions that appeared to be
confirmed, while network partitions or adversarial scheduling can delay
inclusion and finality~\cite{garay2015bitcoin,gervais2016security,natoli2016blockchain}.
More generally, partition tolerance forces distributed services to trade
availability against strong consistency~\cite{gilbert2002brewer}. Consequently,
users who are offline---or connected through intermittent, high-latency, or
low-bandwidth links---cannot independently establish the latest ownership
state, submit a transfer, or know that competing spends have been excluded.
Consensus therefore provides global coherence, but it also places every payment
behind a communication and collective-acceptance boundary.

Several lines of work reduce this cost. Payment channels and the Lightning
Network move repeated transfers off chain~\cite{poon2016bitcoin}; sidechains,
commit chains, and optimistic or zero-knowledge rollups batch execution and
publish compact commitments to a base layer~\cite{kalodner2018arbitrum}; and
protocol-level techniques such as sharding and alternative consensus designs
seek greater throughput and lower confirmation latency~\cite{croman2016scaling}.
These mechanisms are valuable scaling tools, but no well-established approach
eliminates dependence on shared state while retaining robust
double-spend prevention: channels require setup and settlement, rollups inherit
base-layer availability and finality, and offline recipients cannot in general
determine whether the same asset has been spent elsewhere. Meanwhile, advances
in trusted execution environments and secure hardware have enabled protected
execution and attestable state even on untrusted hosts
~\cite{costan2016intel,cheng2019ekiden}.
Open and modular architectures such as Keystone now make TEE mechanisms easier
to inspect, adapt, and deploy beyond a single proprietary enclave design
~\cite{lee2020keystone}. Recent systematization also shows rapid progress in
attestation, isolation, and enclave interfaces across architectures
~\cite{li2024sok}. Together, these developments make the IVM's core abstraction
increasingly plausible, although secure persistent state, rollback resistance,
physical attacks, and lifecycle management remain active research problems.
Motivated by these advances, we ask
whether monetary validity itself can be made locally verifiable. We explore an
\emph{Intrinsic Value Module (IVM)} as a specification for non-copyable,
stateful trusted hardware that allows a coin's authenticity and spendability to
be checked and transferred without contacting an online consensus service.

\paragraph{Contributions.}
We define an IVM abstraction for protected, non-clonable monetary state and
specify an offline transfer protocol. Under explicit assumptions about
attestation, key protection, non-rollback storage, and atomic state
transitions, we establish local transfer atomicity, value conservation, and
resistance to certificate forgery, replay, duplicate credit, and double
spending by a genuine IVM. We do not claim that disconnected IVMs establish a
globally agreed transaction history.

\paragraph{Closest related work.}
Prior monetary systems do not use \emph{intrinsic consensus} for this
abstraction. The closest lines of work are offline electronic cash, which
detects double spending when coins are deposited~\cite{chaum1990untraceable};
payment channels and TEE-backed variants, which retain a base-ledger lifecycle
~\cite{poon2016bitcoin,lind2017teechan}; and trusted-hardware CBDC protocols
supporting temporarily offline, transitive payments~\cite{christodorescu2021twotier}.
The proposal is closest to trusted-hardware offline CBDC designs and TEE-backed
payment systems. Its distinguishing abstraction is that the protected balance
itself is the transferable monetary object, not a temporary representation of
an account or channel position. Table~\ref{tab:comparison} distinguishes these
architectures.

\begin{table}[t]
  \caption{Architectural comparison (\CIRCLE~= full, \LEFTcircle~= partial,
  \Circle~= absent). IVM entries are assumed properties of the abstraction,
  not measured results.}
  \label{tab:comparison}
  \centering
  \scriptsize
  \setlength{\tabcolsep}{4pt}
  \renewcommand{\arraystretch}{1.15}
  \setlength{\abovecaptionskip}{2pt}
  \setlength{\belowcaptionskip}{0pt}
  \begin{tabular}{@{}lcccc@{}}
    \toprule
    Approach & Offline & Re-spend & Open & Intrinsic\footnotemark[1] \\
    \midrule
    Chaum et al.~\cite{chaum1990untraceable}
      & \CIRCLE & \Circle & \CIRCLE & \Circle \\
    Lightning~\cite{poon2016bitcoin}
      & \LEFTcircle & \LEFTcircle & \Circle & \Circle \\
    Teechan~\cite{lind2017teechan}
      & \CIRCLE & \Circle & \Circle & \Circle \\
    Two-tier CBDC~\cite{christodorescu2021twotier}
      & \CIRCLE & \CIRCLE & \LEFTcircle & \LEFTcircle \\
    IVM (this work)
      & \CIRCLE & \CIRCLE & \CIRCLE & \CIRCLE \\
    \bottomrule
  \end{tabular}
  \vspace{-4pt}
  \footnotetext[1]{``Intrinsic'' means not represented by an online account
  after issuance.}
\end{table}

``Offline'' means that payment needs no online authority; direct communication
between payer and recipient is still required. ``Re-spend'' asks whether value
just received can fund a later offline payment. ``Open'' excludes a
pair-specific prefunded channel, while ``intrinsic'' excludes a backing online
account or base-ledger position after issuance. Chaumian cash permits offline
acceptance but normally detects double spending only at deposit. Lightning and
Teechan avoid base-layer interaction during payment, but funds remain channel
positions that eventually settle on chain. The two-tier CBDC design is the
closest predecessor: trusted devices prevent double spending and received funds
can circulate offline, although enrollment, deposit, withdrawal, and redemption
remain tied to server-maintained accounts.


\section{System Model}

We consider participants who control host devices containing an \emph{Intrinsic
Value Module (IVM)}. An IVM executes fixed trusted code and stores persistent
state consisting of a balance, a unique signing key, and protocol metadata. Only
the IVM may change this state; the untrusted host may request operations but
cannot read secret keys or directly set a balance. At manufacture or enrollment,
a trusted root certifies the IVM's public key, implementation, and security
properties. Initial issuance provisions balances through a separately authorized
operation; monetary policy, redemption, and recovery from a lost device are
outside our core model.

We assume standard cryptography and trusted hardware that prevents key
extraction, state cloning, rollback, and modification, even against the device
owner. The adversary may control hosts and communication links, own multiple
genuine IVMs, and observe, delay, drop, reorder, or replay messages. It may not
cause a genuine IVM to deviate from its trusted code or forge another IVM's
attestation. The manufacturer and issuance roots are trusted; root compromise,
hardware side channels, denial of service, and physical destruction are
considered separately. The protocol requires neither a global ledger nor a
wide-area network, clock, or continuously available third party.

\begin{figure*}[htbp]
  \centering
  \includegraphics[width=\textwidth]{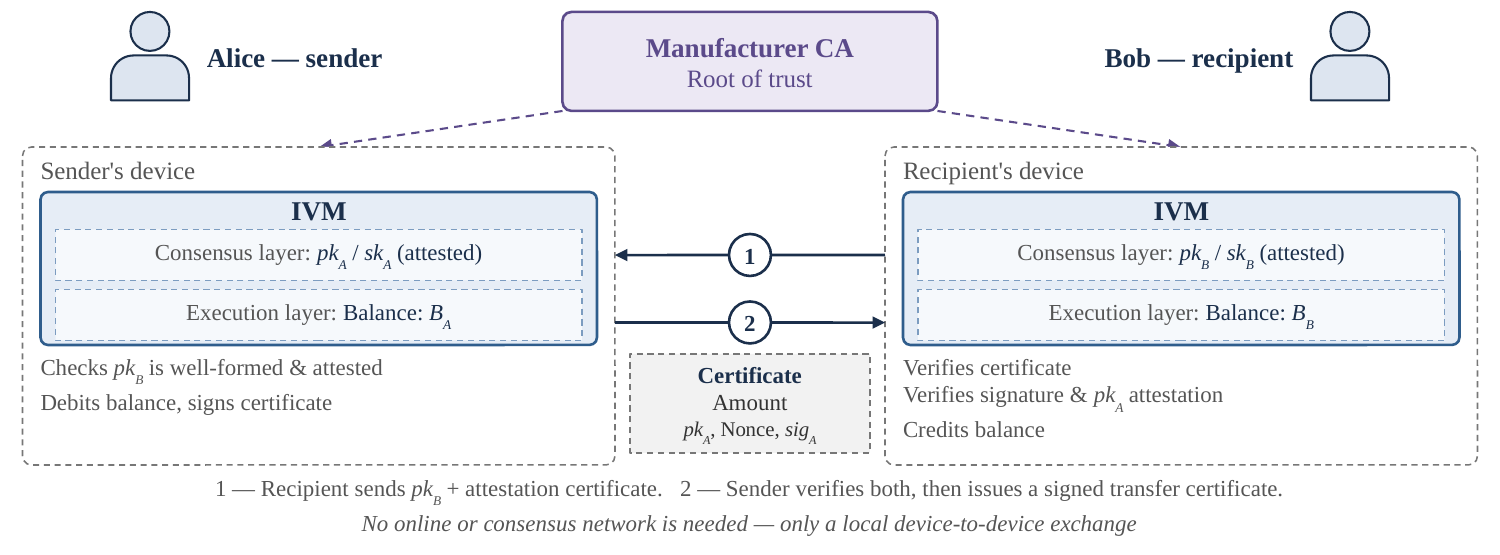}
  \vspace{-10pt}
  \caption{Offline transfer between attested IVMs without a trusted channel,
  ledger, or online authority.}
  \vspace{-10pt}
  \label{fig:ivm-transfer}
\end{figure*}

Figure~\ref{fig:ivm-transfer} shows an offline transfer. Alice's and Bob's IVMs
authenticate one another using their certified keys over an otherwise untrusted
local channel. Bob supplies a fresh transfer identifier; Alice's IVM verifies
Bob's attestation, irreversibly debits the amount, and signs a recipient-bound
transfer certificate. Bob's IVM verifies and records that certificate before
crediting the amount, accepting each identifier at most once. The required
safety properties are conservation of value, authorization, and resistance to
forgery, replay, and double spending. Liveness requires only eventual delivery
between two honest, functioning IVMs; interruption after debit may delay or
destroy value unless a recovery mechanism is added, but must never duplicate it.

\section{Consensus and Execution Layers}

Traditional digital money represents ownership in a replicated history: a payment is valid when the consensus system places it after the payer's credits and before any conflicting spend. Our proposal splits this role into two layers inside the asset-holding hardware. A consensus layer accepts only transitions that preserve non-clonable state and protocol invariants, so that value controlled by one IVM cannot authorize two outgoing transfers; we call this property \emph{intrinsic consensus}, since two genuine IVMs can determine the validity and result of a transfer from authenticated local state alone. A separate execution layer, invoked only once the consensus layer commits, applies the actual state transition --- a balance debit or credit in this protocol, though the same guarantee could authorize richer transition functions.

\paragraph{Meaning of intrinsic consensus.}
In this paper, intrinsic consensus does not mean agreement among independent network participants on a global transaction history. It denotes local agreement between cooperating IVMs about whether one authenticated state transition may commit. The resulting guarantee is local transfer atomicity and asset conservation under the hardware assumptions. It does not provide global state consistency across disconnected devices that have not interacted. We distinguish three properties:
\begin{itemize}
  \item \textbf{Local transfer atomicity}: a sender-recipient session commits at most once.
  \item \textbf{Asset uniqueness}: a balance cannot authorize two outgoing transfers.
  \item \textbf{Global state consistency}: all devices agree on one transfer history.
\end{itemize}
The protocol claims the first two under its assumptions and does not independently establish the third during disconnection.

\subsection{Consensus-Layer and Execution-Layer State}

The consensus layer stores certified identity, a signing key, and any unfinished transfer session; a certificate binds this identity to a hardware version and trusted program measurement under a manufacturer root, and peers are accepted only if their certification chain and program version satisfy a common policy. The execution layer stores the unit's spendable state --- a balance in this protocol --- and exposes only transitions the consensus layer has authorized. Issuance is a distinguished transition, authorized by an issuer credential and recorded by the receiving IVM; ordinary transfers can only move existing units and cannot invoke issuance or increase the combined balance of the participating modules.

The execution layer implements a single fungible balance, which suffices for payments and keeps the analysis simple. Nothing in the consensus layer depends on this choice: it guarantees only that an authorized transition is applied at most once, regardless of what that transition computes. An execution layer could therefore expose multiple asset types, programmable spending conditions, or other deterministic transition logic. We leave such designs to future work and restrict the security analysis to the balance case.

\subsection{Offline Transfer}

A transfer of amount $v$ from Alice to Bob proceeds as follows.

\begin{enumerate}
  \item \textbf{Request.} Bob's IVM creates and persistently reserves a fresh
  session identifier $n$, then sends Alice an attested request
  $(\mathit{pk}_B,n,v)$. At most one transfer may consume this reservation.
  \item \textbf{Debit.} Alice's IVM verifies the request and that its available
  balance is at least $v$. It atomically subtracts $v$, records a pending
  transfer, and signs a certificate binding $v$ to $(\mathit{pk}_B,n)$.
  \item \textbf{Credit.} Bob's IVM accepts the certificate only if the identity,
  amount, and reserved session match. It atomically marks $n$ consumed, adds
  $v$ to its balance, and returns a signed receipt.
  \item \textbf{Completion.} Alice's IVM verifies the receipt and deletes its
  pending record. Retransmission is idempotent: replaying a request, certificate,
  or receipt cannot produce another debit or credit.
\end{enumerate}

After Alice commits the debit, the amount is either credited to Bob's reserved
session or locked in Alice's pending state; no path restores it while the
certificate may still be accepted. The progression is
$\mathit{Available}\rightarrow\mathit{Pending}(n,\mathit{Bob},v)\rightarrow\mathit{Completed}$,
with $\mathit{LockedAfterFailure}$ reachable only from $\mathit{Pending}$.

Both modules commit every security-relevant update to non-volatile protected
state before releasing the corresponding message. A host may interrupt or
repeat any step, but can only postpone completion. If the certificate or receipt
is lost, the recorded session permits retransmission when the devices reconnect.
If Bob's IVM is permanently destroyed after Alice commits the debit, the amount
remains locked; reclaiming it safely requires an additional recovery authority
or timeout mechanism and is not part of the base protocol.

\subsection{Local Verification and Finality}

Bob accepts value based only on Alice's certificate, both attestation chains,
and his own protected session state. Acceptance is immediate after the local
exchange and is not probabilistic: there are no blocks to await and no later
chain reorganization. The protocol nevertheless replaces, rather than removes,
trust: ledger consensus is replaced by assumptions about correct hardware,
attestation roots, and non-rollback storage. This local finality says nothing
about availability: either party can abort before credit, and revocation or
software upgrades cannot be learned offline unless their validity is encoded in
credentials with an appropriate lifetime policy.


\section{Security Analysis}

We assume unforgeable signatures, sound attestation, and IVM state that cannot
be extracted, cloned, modified, or rolled back. Let $S$ be the total value
introduced by authorized issuance and $B_i$ the spendable balance held by IVM
$i$'s execution layer.

\paragraph{Conservation and double spending.}
The protocol maintains $\sum_i B_i \leq S$. Before producing a certificate,
Alice's IVM atomically reduces $B_A$ and records a unique pending session; it
never restores that amount merely because a message or receipt is missing.
Thus, the same balance cannot back two certificates for genuine, uncompromised
IVMs under the stated assumptions. Bob credits a certificate only when its
recipient key and reserved identifier match, and atomically consumes that
identifier. The certificate therefore cannot increase one or several spendable
balances more than once. Equality may be temporarily lost while value is
pending, but the supply cannot increase.

\paragraph{Authenticity, authorization, and replay.}
Every accepted certificate is signed by an attested sender key and binds its
amount, recipient, and fresh session identifier. Alteration or redirection
invalidates the signature. Replaying a request returns the existing pending
transfer instead of causing another debit; replaying a certificate encounters
a consumed recipient session; and replaying a receipt only completes deletion
of Alice's existing pending record. A hostile host or channel can delay or
suppress a payment, but cannot forge or duplicate one under the stated
assumptions.

\paragraph{Finality and limitations.}
Once Bob records the credit in protected state, the payment is locally final:
there is no competing chain or later reorganization. A lost receipt leaves
Alice's non-spendable pending record intact, while retransmission lets Bob
return the same receipt without crediting again. Permanent loss of Bob's module
may lock the value; unilateral recovery is unsafe without proof that Bob did
not and can no longer accept the certificate. These guarantees collapse if an
attestation or issuance root is compromised, a key is extracted, or protected
state is cloned or rolled back. Availability, physical destruction, and
side-channel resistance are not guaranteed by the protocol.


\subsection{What the protocol does not solve}

The protocol provides no global transfer ordering, no offline revocation, no
safe unilateral recovery after ambiguous debit, no protection after root
compromise, no availability under device destruction or communication failure,
no security against colluding devices beyond the stated model, and no empirical
claim that current hardware meets these assumptions.

\section{Discussion and Conclusion}

Intrinsic consensus changes the location and shape of the trusted computing
base. It removes miners, validators, global transaction propagation, and ledger
storage from the critical path of an ordinary payment, allowing constant-size
local communication and operation during network disconnection. This makes the
model useful for disaster settings, remote regions, low-value machine-to-machine
payments, and other environments in which continuous access to a base ledger is
undesirable. It may also complement rather than replace an online currency:
users could provision IVM balances online and later exchange them offline.

The cost is strong trust in a hardware ecosystem. A scalable deployment must
address manufacturing faults, root compromise, device revocation, secure
updates, interoperability among vendors, privacy, lost or destroyed modules,
and auditable issuance. Some goals are in tension with offline operation:
immediate dissemination of revocations and safe recovery after ambiguous
failures normally require external state. Side-channel and fault-injection
resistance also remain difficult in existing trusted hardware, so the IVM
should be read as a requirements specification rather than a claim that present
devices already provide the full abstraction.

Can digital money enforce uniqueness within the object carrying value rather
than through global agreement? The IVM combines attested identity,
non-extractable keys, non-clonable state, atomic balance transitions, and
one-time transfer sessions to enable locally final, value-conserving offline
payments. Future work must validate these assumptions and provide recovery
without restoring continuous consensus.


\bibliographystyle{ACM-Reference-Format}
\bibliography{references}

\end{document}